\documentclass[times,3p,final]{elsarticle}

\usepackage{graphicx}
\usepackage{amssymb}
\usepackage{amsmath}
\usepackage{epsf}

\usepackage{ulem}
\usepackage{color}

\journal{Physica A}

\begin{document}

\begin{frontmatter}

\title{A spring-block analogy for the dynamics of stock indexes}

\address[label1]{Institut f\"ur Theoretische Physik, Goethe Universit\"at, Frankfurt am Main, Germany}
\address[label2]{Babe\c{s}-Bolyai University, Department of Physics, Cluj-Napoca, Romania }
\address[label3]{Edutus College, Tatab\'anya, Hungary}
\cortext[cor1]{Corresponding author.}

\author[label1,label3]{Bulcs\'u S\'andor} 
\author[label2,label3]{Zolt\'an N\'eda\corref{cor1}}\ead{zneda@phys.ubbcluj.ro}

\begin{abstract}
A  spring-block chain placed on a running conveyor belt is considered for modeling stylized facts 
observed in the dynamics of stock indexes. Individual stocks are modeled by the blocks, 
while the stock-stock correlations are introduced via simple elastic forces acting in the springs. 
The dragging effect of the moving belt corresponds to the expected economic growth. 
The spring-block system produces collective behavior and avalanche like phenomena, similar to the ones 
observed in stock markets. An artificial index is defined for the spring-block chain, and 
its dynamics is compared with the one measured for the Dow Jones Industrial Average. 
For certain parameter regions the model reproduces qualitatively well the dynamics of the logarithmic 
index, the logarithmic returns, the distribution of the logarithmic returns, the avalanche-size 
distribution and the distribution of the investment horizons. A noticeable success of the model is 
that it is able to account for the gain-loss asymmetry observed in the inverse statistics. 
Our approach has mainly a pedagogical value, bridging between a complex socio-economic phenomena 
and a basic (mechanical) model in physics.  

\end{abstract}


\end{frontmatter}

\section{Introduction}

Spring-block (SB)  models have been used for a long time to model complex phenomena in physics 
and engineering. This model family was introduced by Burridge and Knopoff \cite{BK} for describing 
the distribution of earthquakes after their magnitudes. In its original 
version a one-dimensional chain of blocks connected by springs is placed on a  moving plane. 
The blocks are free to slide on this plane, subject to a velocity dependent friction force. 
All blocks are connected with additional springs to a second plane, which
is in rest, and it is placed above the spring-block chain. 
This system was meant to describe two tectonic plates that are in 
relative motion respective to each other, exhibiting a complex stick-slip dynamics. 
In such an approach the slipping motion of the blocks will lead to energy dissipation and 
the complex avalanche-like dynamics will yield a scale-free distribution for the energy dissipated 
in avalanches. The model exhibits a complex dynamics and 
Self-Organized Criticality (SOC) \cite{Bak1,Carlson}.

This very simple physical system formed by an ensemble of blocks interconnected with springs and placed 
on a frictional surface resulted in many interesting applications. It was successful in reproducing 
desiccation patterns and dynamics of crack formation in mud, clay or thin layers of paint 
\cite{Andersen,Leung}, self-organized patterns in wetted nano-sphere  \cite{Jarai1} or 
nano-tube \cite{Jarai2} arrays or even crack structures obtained in glass \cite{Horvat}. 
It was applied for describing the Portevin Le Chatelier effect \cite{Lebo1} and magnetization phenomena 
in ferromagnets, including the Barkhausen noise \cite{Bark1}. Besides applications in physics, 
the spring-block model got some interdisciplinary applications as well. Formation of traffic jams 
in a single-lane highway traffic \cite{Jarai3} or detection of region-like structures \cite{Mate} 
in a delimited geographic space are a few recent applications in such sense. 
Continuing this line of studies, here we intend to consider a simple one-dimension spring-block chain 
for revealing a pedagogically useful and interesting analogy with the dynamics of stock indexes.

\begin{figure}[ht!]
  \centering
  \begin{center}
    \includegraphics[scale=0.6]{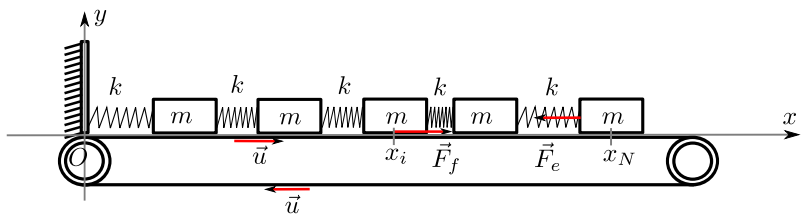}
  \end{center}
  \caption{ The  used spring-block system. Blocks of mass $m$ connected by springs with spring 
	   constant $k$ are placed on a conveyor belt that is moving with a constant velocity $u$.}
  \label{fig1}
\end{figure}

The simplest version of the model will be considered here, 
the one referred in the literature as "train model" \cite{Sousa}. 
As it is sketched in Figure \ref{fig1}, a spring-block chain is placed on a running conveyor-belt, 
so that the first block is fixed with a spring to an external, static point. As a result of the 
dragging effect of the conveyor belt, the chain is stretched and a complex-stick slip dynamics emerges. 
Both the case of one block alone \cite{Sousa2,Baumberger,Vasconcelos,Elmer,Sakaguchi,Johansen} 
and the case of a chain formed by several blocks  
\cite{Sousa2,Vasconcelos,Sousa4,Chianca,Sousa5,Szkutnik,Erickson,Bulcsu} were considered in previous
theoretical and experimental studies. Coexistence of chaotic dynamics and SOC was observed by 
many authors \cite{Sousa2,Bulcsu}. Nonlinearity was introduced in the model via friction forces. 
Several friction force profiles were considered, starting from velocity-weakening friction forces 
combined with a constant static friction force \cite{Sousa2,Sousa5,Erickson} to simple state-dependent 
friction forces \cite{Sakaguchi,Bulcsu}.  Our aim here is rather different 
from these previous studies. Instead of mapping the dynamical complexity of such 
a system, we take a different turn and use the system as a simple analogy for modeling the dynamics 
of stock indexes. 

The dynamics of stock indexes are in the focus of physicists from a  quite long time 
\cite{Mantegna}. The existence of many {\em stylized facts} in the financial market 
(see for example \cite{Mantegna}) captured the interest of the statistical physics community. 
Many simple models have been used to reproduce statistical features of price/index fluctuations 
(for a review see for example \cite{Sinha}). Definitely, the most basic approach among them 
 is the simple random walk (or Brownian dynamics) model applied to the logarithmic index \cite{Bouchaud}. 
This model is known as the geometric random walk model \cite{Malkiel}. The fact that dynamics 
of stock prices can be approached by a simple geometric random walk is one of the 
most interesting empirical fact about financial markets . This simple model was first proposed 
by the French mathematician Louis Bachelier in the early 1900, and it has been strongly debated 
since then.  The most important support for this model comes from the experimental fact that 
volatility of stock returns tend to be approximately constant in long term. 
Although this model cannot account of many important statistical aspects of the index or 
stock price fluctuations (such as time-varying volatility \cite{Schwert,Patzelt}, 
evidence of some positive autocorrelations \cite{Iwaisako}, or the asymmetry of the investment 
horizons distribution for positive and negative return levels \cite{Simonsen1}), 
it' s simplicity and intuitive nature makes it pedagogically useful.  It can be considered 
as a first step (zero order model) towards understanding the nature of the stock index dynamics 
by a general model of mathematics and physics. 

Similarly with random walk, the SB system is also a general model of physics, 
which is appropriate for capturing in an elegant manner universal trends in the dynamics 
of stock indexes. Although the physical picture behind the two phenomena 
(motion of a spring-block chain and the dynamics of stock indexes) is rather different, 
a simple and useful analogy can be drawn between them.
This analogy might be useful for pedagogical reasons and for understanding some stylized facts. 
Our motivation here was not to give a model which performs better than the nowadays used 
rather complex approaches \cite{Comte}.  Instead of this we focus on the simplicity 
and visuality of the model, offering a pedagogical picture that is one step 
ahead of the basic random walk approach. The rest of the paper is about the proposed 
analogy between the dynamics of stock indexes and the simple spring-block chain system, 
and also about discussing the modeling power of such an approach.

\section{Stylized facts for the dynamics of stock indexes}

The performance of stocks and markets, and thus the performance of the economy of a region or state, 
over a certain time history is traditionally measured by the distribution of the  
$r_{\Delta t}(t)$ logarithmic return \cite{Simonsen1},  which gives the generated return over 
a certain $\Delta t$ time period. For stock market indexes it is defined as the 
natural logarithm of the price change over a fixed $\Delta t$ time interval:

\begin{equation}
 r_{\Delta t}(t)=s(t+\Delta t)-s(t)=\ln{\frac{S(t+\Delta t)}{S(t)}}
\end{equation}  

where $s(t)=\ln(S(t))$ is the logarithmic index ($S(t)$ is the value of the index or 
the price of a stock). The logarithmic return indicates, how much the logarithmic index or 
stock prices have increased or decreased in a fixed time interval, thus presents 
the economical performance of a company. Among others, the advantage of using logarithmic returns is, 
that it is an additive quantity.

The financial time series considered as  example in the followings is the Dow Jones Industrial Average 
(DJIA) index. We deal with the daily closure prices for about an 80 year long time period 
(more than 20000 trading days), from 1.10.1928. to 1.02.2011. We have chosen the DJIA index, 
because this is the oldest continuously functioning index and it can be freely downloaded 
from the internet. It is an index that shows how 30 large publicly owned companies 
(like General Electric, IBM, Microsoft, McDonald's, Coca-Cola, etc.)  based in the United States 
have traded during a standard trading session in the stock market. The value of the DJIA index 
can be calculated quite simply, though is not the actual average of the prices of its component stocks, 
but rather the sum of the component prices divided by a divisor, which changes whenever one of 
the component stocks has a stock split or stock dividend, so as to generate a consistent value 
for the index \cite{Sullivan}.

\begin{figure}[ht!]
  \centering
  \begin{center}
    \includegraphics[scale=0.35]{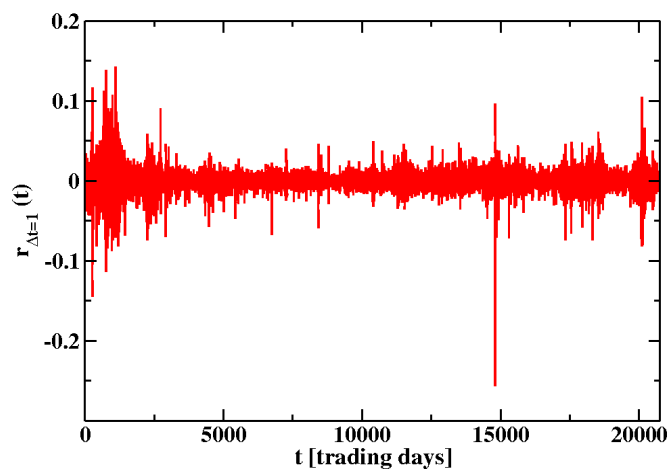}
  \end{center}
  \caption{ Daily logarithmic returns of the DJIA index as a function of trading days 
	   from 1.10.1928. to 1.02.2011. Large negative values indicate events 
	   when the market crashed. }
  \label{fig2}
\end{figure}

As an example in Fig. \ref{fig2}. the time series of daily 
returns for DJIA are shown (this means we consider $\Delta t=1$ day long time intervals). 

The standard deviation of $r_1(t)$ daily log-returns is called \textit{volatility}. 
The $\sigma$ volatility is a measure for the variation of price of a financial 
instrument (e.q. stocks), therefore in the case of higher volatilities there is a 
higher probability of large price fluctuations, thus for large gains or losses. 
For the DJIA index, the historical volatility of the daily log-returns 
is about $\sigma=0.011$, i.e. $\sigma\approx 1\%$. 

The distribution of logarithmic returns measure how much an initial investment, made at time $t$, 
has gained or lost by the time $t+\Delta t$.  Empirical results have demonstrated, 
that the distribution of returns can be approximated by a Gaussian distribution, 
although there are meaningful differences, such as the presence of fat tails 
\cite{Mantegna,Bouchaud,Simonsen1,Mantegna2}.  The fat tails correspond to a much larger 
probability for large price changes than what to be expected from a Gaussian statistics, 
an assumption made in the mainstream theoretical finance \cite{Mantegna,Bouchaud,Jensen}.

The method of inverse statistics in economics was recently suggested by Simonsen et al.
\cite{Simonsen1}, being inspired by earlier works in turbulence \cite{Jensen2}. 
The method was adopted as an alternative measure of the financial market performance.
The idea is to turn the problem around and ask the inverse question: 
what is the typical waiting time to generate a fluctuation of a given size 
in the price \cite{Simonsen1}? For this we have to determine for an index 
or a stock the distribution of $\tau_{\rho}$ time intervals needed to obtain 
a predefined $\rho$ return level. We search thus for the shortest $\Delta t$ interval, 
for which it is true, that:

\begin{equation}
 r_{\Delta t}(t)\geqslant \rho, \text{ if  } \rho>0,
\end{equation} 

or

\begin{equation}
 r_{\Delta t}(t)\leqslant \rho, \text{ if  } \rho<0.
\end{equation} 

Practically if given a fixed $\rho$ logarithmic return barrier (proposed by the investor) 
of a stock or an index, as well as a fixed investment date (when the investor buys some assets), 
the corresponding time span is estimated for which the log-return of the stock or index 
for the first time reaches the level $\rho$. This is also called \textit{first passage time} 
through the level $\rho$ \cite{Simonsen1}. In a mathematical formulation (for $\rho>0$) 
this is equivalent with:

\begin{equation}
 \tau_{\rho}=\inf\{\Delta t\geqslant 0\,\,|\,\, r_{\Delta t}(t)\geqslant \rho\}.
\end{equation} 

\begin{figure}[ht!]
  \centering
  \begin{center}
    \includegraphics[scale=0.7]{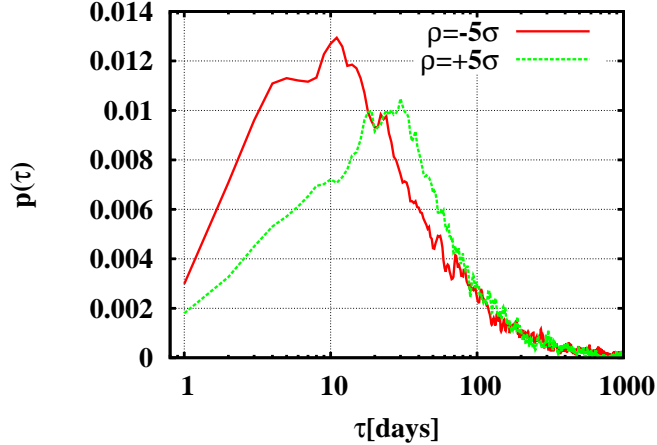}
  \end{center}
  \caption{The investment horizon distributions for the DJIA closing prices for positive and negative 
	   return levels. The $p_{\rho}(\tau)$ probability distribution function is measured 
	   in $\tau$ trading days, at the standard level of $|\rho|=5\sigma\approx 0.05$ 
	   (i.e. 5\% return) \cite{Simonsen1,Siven}.  For positive returns we notice the maximum 
	   of this function at approximately $\tau^{\star}_{\rho}=25$ trading days, which is 
	   the most probable time of producing a return of minimum 5\%.   For both values of the return,
	   the probability distribution functions are characterized by a single maximum, 
	   though there is shift in the optimal investment horizons, which is termed 
	   in the literature as the gain-loss asymmetry \cite{Jensen}.}
  \label{fig3}
\end{figure}

In the literature this $\tau_{\rho}$ time is termed as the \textit{investment horizon} 
for the proposed $\rho$ log-return for that stock or index \cite{Simonsen1}. 
The investment horizon indicates for an investor the $\tau_{\rho}$ time interval he has to wait, 
if the investment was made at time $t$, to achieve the proposed $\rho$ (e.q. $5\%$) log-return 
at $t+\tau_{\rho}$. The normalized histogram of the accumulated values of the first passage times 
form the $p_{\rho}(\tau)$ probability distribution function of the investment horizons. 
The method described above is called the method of inverse statistics.
The distribution of investment horizons for the DJIA index is presented in Fig. \ref{fig3}.
The maximum of the distribution function determines the most probable waiting time for that log-return, 
or in other words the most probable investment horizon. This is the \textit{optimal investment horizon} 
for that stock or index. The first passage distribution gives also information about the stochasticity 
of the underlying asset price \cite{Biferale,Jensen2}.

Building up the inverse statistics of DJIA index also for negative return levels (i.e. $\rho=-5\%$), 
it was found \cite{Jensen,Simonsen3} that the distribution of investment horizons is similar 
to the one for positive levels (with a pronounced maximum), though there is one important difference. 
For negative return levels the maximum of the probability distribution function is shifted to left, 
generating about a $\Delta \tau^{\star}_{\rho}\approx 15$ trading days difference in the 
optimal investment horizons. In Fig. \ref{fig3} this asymmetry of the inverse statistics 
is presented for $\rho=+5\sigma$ and $\rho=-5\sigma$ log-returns.
Later it was found, that the asymmetry of inverse statistics is present for every 
important stock index, thus stock markets present a universal feature, called the 
\textit{gain-loss asymmetry} \cite{Jensen}. In contrary with the indexes, stock prices 
show a smaller degree of asymmetry \cite{Siven,Balogh}.

Minimal models have been proposed for explaining this apparent paradox. 
One of these models is termed the fear-factor model \cite{Donangelo,Simonsen3}. 
In this model a synchronization-like concept is introduced, the so-called fear factor. 
The presence of this fear factor at certain times causes the stocks to all move downward, 
while at other times they move independently from each other. 
In this way the fear-factor model assumes stronger stock-stock correlations during dropping markets 
than during market raises \cite{Balogh}. 
Recently the idea of fear factor model was generalized by allowing longer time periods 
of stock-stock correlations \cite{Siven}. Balogh et. al. \cite{Balogh} have demonstrated 
by conducting a set of statistical investigations on the DJIA index and its constituting stocks, 
that indeed there is stronger stock-stock correlation during falling markets. 
This empirical result gives confidence in the fear factor hypothesis.

An additional explanation for the gain-loss asymmetry is given by a simple 
one factor model \cite{Bouchaud2} and the Frustration Governed Market model \cite{Ahlgren}, 
which incorporates the leverage effect. Very recently the problem has also been investigated 
by the group of D. Sornette \cite{Lagger}. 

\section{The spring-block analogy}

We consider the simple one dimensional SB system  
sketched in Fig. \ref{fig1}, referred in the literature as ``train model'' \cite{Sousa}. 
A spring-block chain, composed of $N$ identical blocks, 
is placed on a platform (conveyor belt) that moves with constant velocity. 
The first block is connected by a spring to a static external point.   As a result, 
due to the dragging effect of the moving platform, the blocks will undergo a complex 
stick-slip dynamics \cite{Bulcsu}.

As the conveyor belt is started with $u$ velocity, the whole system moves together with the belt 
until the first block starts to slip.  The slipping moment occurs when the sum of elastic forces 
acting on a block overcomes the maximal value of the static friction force ($F_{st}$). 
The slipping motion is tempered by the $F_f$ dynamical friction forces. 
The block sticks again to the belt when its relative velocity $v_r=v_i - u$ becomes zero. 
After that the process may start again 
(a movie about the dynamics in the spring-block system can be consulted at \cite{movie}).

In the analogy taken here, the price of individual stocks can be modeled as the $x_i(t)$ positions 
of blocks, measured from an external static point. For example, if we would like to use 
this analogy for the DJIA index that represents the economical state of 30 large American stocks,  
a chain composed of $N=30$ blocks will be considered in the model. 
As one can set up a ranking between the stocks as a function of their price in the index, 
the pre-existing order between the blocks of the chain is not completely irrelevant. 
Of course, the ranking of stocks by prices can change throughout the time. 
The corresponding event would be, that blocks come through each other, 
but this unphysical phenomena is not considered in this simple model.

In the framework of our model the value of the $X(t)$ index is defined as the 
center of mass of the system, measured in the static coordinate system:

\begin{equation}
  X(t)=\frac{\sum_{i=1}^{N}x_i(t)}{N}.
\end{equation}

The elastic forces acting between blocks via springs correspond to  stock-stock correlations.
The conveyor belt acts on the system through the friction forces. 
The pulling imposed by the belt models the desire for grow, a general trend expected 
from all companies and societies. For small belt velocities the stick-slip dynamics, 
will lead to  "avalanches" of different sizes. Therefore, the length of the chain, 
defined by the position of the last block $x_N$ fluctuates as a function of time 
with largely varying amplitudes, as shown in Fig. 8.a. in Ref. \cite{Bulcsu}.
This self organized behavior leads to  abrupt drops in the value of the index, 
when all the stock prices fall in a correlated manner \cite{Balogh}.

In order to account for the random jumps of the index when it is raising continuously 
and also for the observed fat-tail distribution of the positive 
logarithmic returns the $u$ velocity of the belt is randomly varied (more details in Appendix A). 

An important ingredient for making the proposed analogy to work is to choose the sampling time 
in the dynamics of $X(t)$.  The financial time series (the DJIA index in our case) represents 
the daily closure prices for about an 80 year-long time period. 
Similarly with many other studies in the literature, intra-day variability of the index 
was not considered in this study. In the spring-block analogy we need thus to define also 
a discrete time-step, beside the obvious continuous time of the dynamics. 
For this, the time-series of the simulation needs to be sampled with a well defined frequency, 
producing in such manner a discrete time-series that should correspond to a one 
trading day time-interval. This sampling frequency will influence the volatility measured 
in the artificial index generated by the spring-block model.  
Since the volatility of log-returns for the DJIA index is known we can fix the sampling interval 
in the model by imposing  this experimental value (see Appendix A).

The power of the analogy introduced in this model consist in
the fact that the $X(t)$ dynamics shows many similarities with the observed dynamics of stock indexes. 
Details for the dynamics, computer simulations and the used parameters for the spring-block chain 
are given in Appendix A.

\section{Model results in comparison with historical stock market data}

Considering the simulation parameters specified in Appendix A and the integration method 
describer in Appendix B, we compare now the statistical results offered by the 
SB train model and those measured from the DJIA index. The parameter values were chosen 
for reproducing in an optimal manner all the stylized facts measured for the DJIA. 
One can of course choose other parameters that will lead to better results for 
some of the statistical measures, but in such cases larger differences will be obtained for the others. 
We recall however that our stated aim here was not to give a realistic description for the dynamics 
of stock indexes, and it is not surprising thus that with such a simple mechanical analogy 
one cannot reproduce all the observed statistical features for the dynamics of stock indexes. 
 
 \begin{figure}[h!]
  \centering
  \begin{center}
    \includegraphics[scale=0.3]{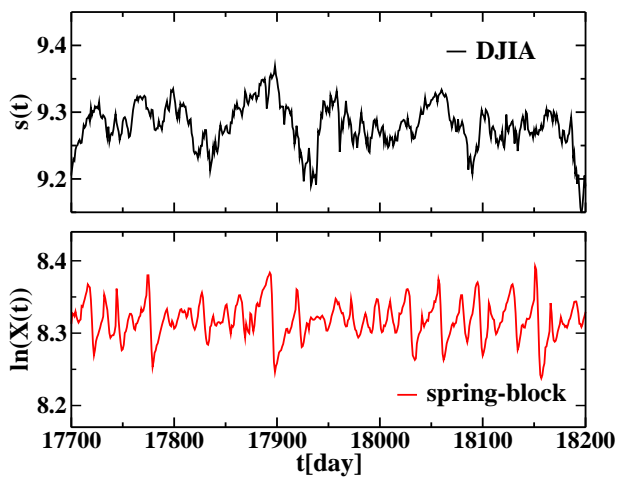}
  \end{center}
  \caption{A sample of the same size (500 days) for the dynamics of the DJIA 
	   logarithmic index and for the logarithmic index  ($\ln[X(t)]$) 
	   derived from the spring-block model.}
  \label{fig4}
\end{figure}

\begin{figure}[h!]
  \centering
  \begin{center}
    \includegraphics[scale=0.3]{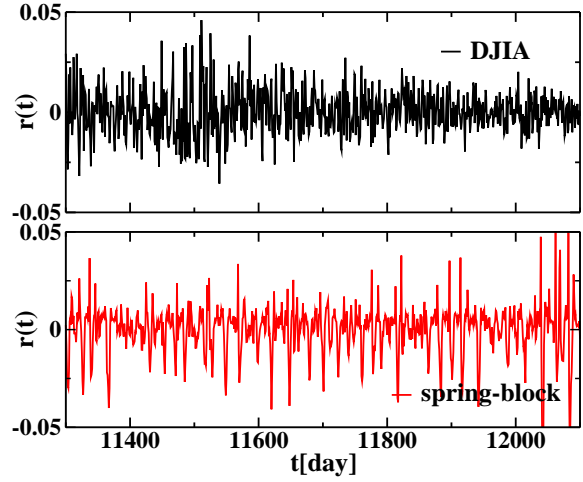}
  \end{center}
  \caption{Visual comparison for a time-period of 800 days between the logarithmic returns 
  (for $\Delta t=1$ day) in the DJIA data and the spring-block model.}
  \label{fig5}
\end{figure}

 In Figure \ref{fig4} we show a first visual comparison between the logarithmic index 
 for the DJIA ($s(t)$) and the one generated by the SB model ($\ln[X(t)]$).  
 In spite of the fact that one can observe visually detectable statistical differences, 
 the results are quite promising. Similar results can be concluded if we analyze 
 the logarithmic returns for $\Delta t =1$ day (Figure \ref{fig5}).

\begin{figure}[ht!]
  \centering
  \begin{center}
    \includegraphics[scale=0.3]{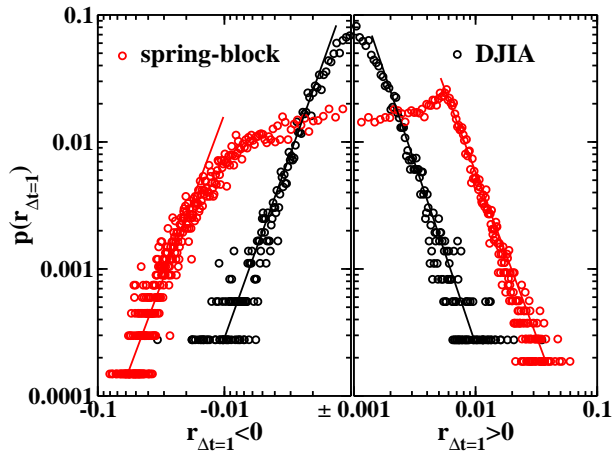}
  \end{center}
  \caption{Distribution of the positive and negative logarithmic returns 
	   ($\Delta t=1$ day) for the DJIA index and the artificial index generated by the SB model.}
  \label{fig6}
\end{figure}

The differences and similarities detected by the visual examination of the logarithmic returns are better illustrated 
if we plot the distribution function of the logarithmic index (Figure \ref{fig6}). From the log-log plot one will conclude that 
the power-law tails observed for high return values are
successfully reproduced by the SB model with the chosen random driving of the conveyor belt. 
There are however significant differences in the low returns limit. Seemingly the
distribution function for the DJIA is more narrow and less asymmetric than the one obtained in our SB approach. 

\begin{figure}[h!]
  \centering
  \begin{center}
    \includegraphics[scale=0.3]{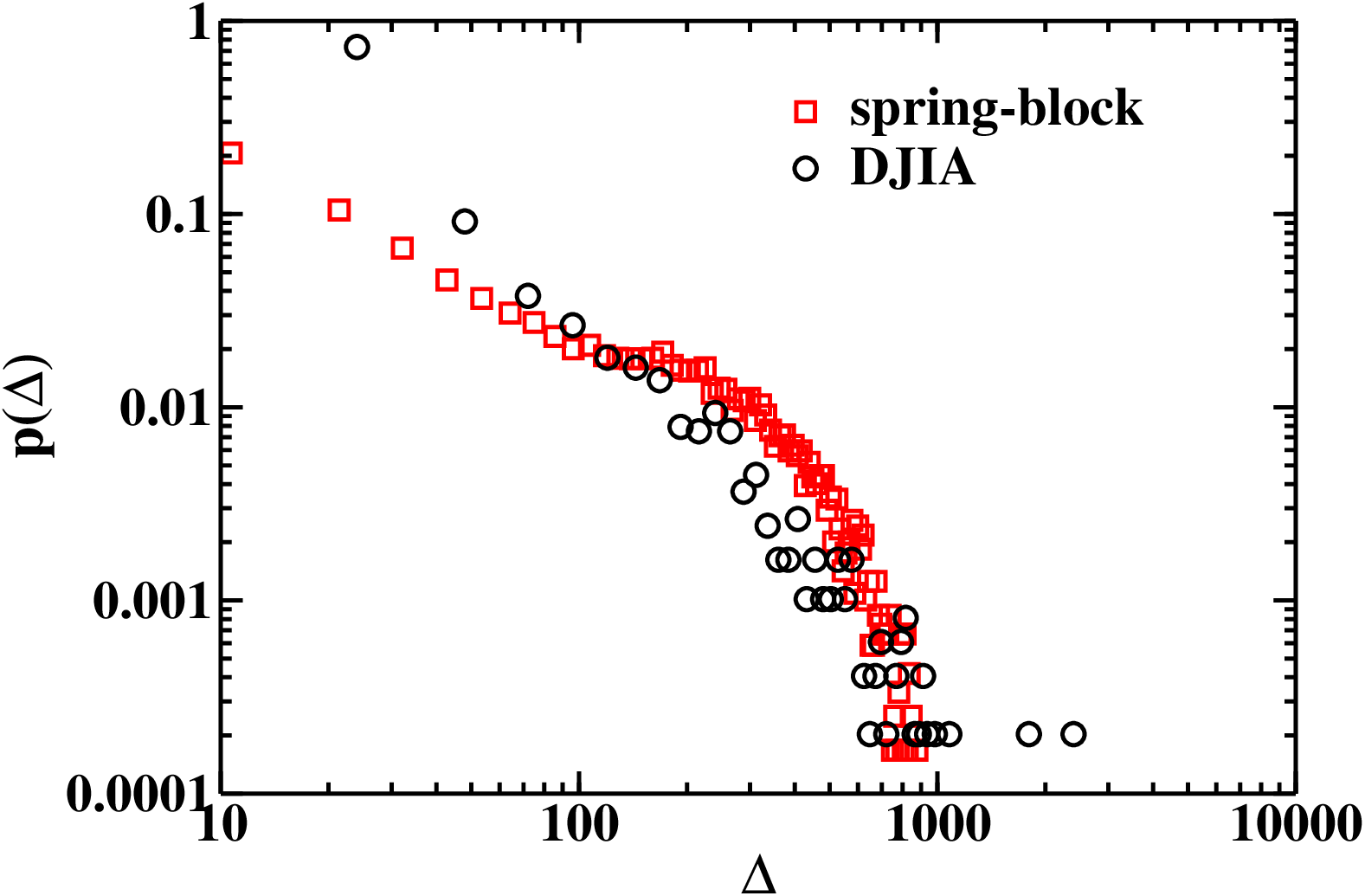}
  \end{center}
  \caption{Avalanche-size distribution in the dynamics of the DJIA index and in the spring-block model.}
  \label{fig7}
\end{figure}

For further statistical comparison one can construct the avalanche-size distribution for both the 
DJIA and the SB model.  An avalanche is defined as a consecutive, monotonic  drop in the index,
between time moment $t_1$ and $t_2$. The distribution of the variation in the index for the avalanches
 ($\Delta=S(t_2)-S(t_1)$ for the DJIA and $\Delta=X(t_2)-X(t_1)$ for the SB model)
generates the avalanche-size distribution. Measured results are compared in Figure \ref{fig7}. 
As it is expectable in both systems there are many small avalanches and one will find also 
a few extremely large ones. This situation is characteristic  for self-organized criticality, 
a feature that seemingly both system exhibit. The two distribution functions are rather similar, 
although the trend for the SB model is more complex than the one obtained for the case of the DJIA.

\begin{figure}[ht!]
  \centering
  \begin{center}
    \includegraphics[scale=0.3]{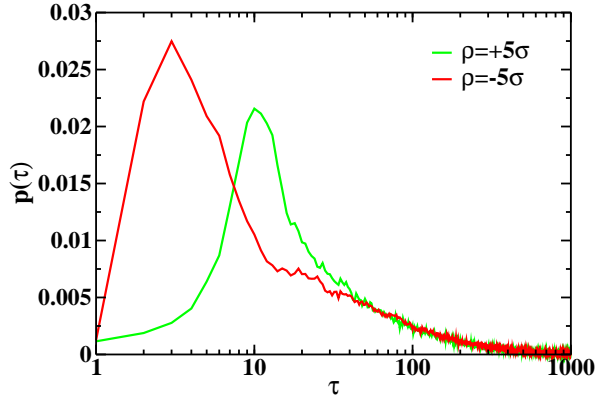}
  \end{center}
  \caption{Distribution of the investment horizon for the artificial index generated by the spring-block model,
   considering $|\rho|=5\sigma\approx 0.05$ (i.e. 5\% return). Results to be compared with the ones
   obtained for the DJIA index (Figure \ref{fig3}). }
  \label{fig8}
\end{figure}

Finally we consider the inverse statistics, i.e. the distribution of the investment horizons for a fixed logarithmic return
value. We considered the same $\rho=+5\sigma$ and $\rho=-5\sigma$ log-return values as in the case 
of the DJIA index (Figure \ref{fig3}) and the results obtained for the SB model are plotted in Figure \ref{fig8}. 
The SB model reproduces with success the gain-loss asymmetry observed in real stock index data. Although the 
position of the maximum for the curves is shifted by a few days, the shape of the curves and
asymmetries in the investment horizons are similar. The SB model is capable thus for reveling also this subtle 
statistical feature characteristic for stock indexes. 

\section{Conclusions}

A simple mechanical analogy has been considered for modeling the  
dynamical features of stock indexes. A chain of blocks connected by springs subjected to the continuous driving of a 
running conveyor belt was used as a model system for stock indexes. 
Contrarily with the large number of recently reported models, here our aim 
was rather different.  Instead of considering  an economically relevant and complex description, we tested the applicability 
of a general and simple model of physics: the spring-block system.  The analogy we have made is well motivated
and highlights the connections among individual stocks prices and their collective behavior.  
We have fixed the driving (speed of the conveyor belt) and the free parameters of the model so 
that most of the stylized facts known for the DJIA stock index are reasonable well reproduced. When judging the
success or fail of our approach one has to take into account also the fact, that we did not focused only on a few
and selected stylized facts, as most of the models proposed in the recent literature does. Instead, we considered all the
used and measurable statistical features of the DJIA stock index, trying to reproduce them as a whole
 with a fixed parameter set. 
An important results of the introduced model is that it accounts for the puzzling phenomenon of gain-loss asymmetry which is observed in the inverse statics, and the model 
describes also well the avalanche-size distribution curves. In spite of the success in
reproducing qualitatively the known statistical features of the index dynamics, a finer statistics
revealed noticeable differences.  We consider this normal, since one cannot expect from a such a simple mechanical analogy a deeper understanding for a complex socio-economic phenomenon. 

In conclusion, we believe that the model introduced here has mainly a 
pedagogical value for the interdisciplinary field of econophysics. It offers an appealing mechanical analogy for
physicist to approach collective phenomena in stock markets.
 
\section*{Acknowledgments}

This research was supported by the European Union and the State of Hungary,
co-financed by the European Social Fund in the framework of T\'AMOP 4.2.4.A/2-11-1-2012-0001 
‘National Excellence Program’.

\section*{Appendix A. Details of the dynamics for the spring-block system}

We consider a chain formed by $N$ identical blocks of mass $m$. The blocks are connected by identical springs of linear spring constant $k$ and undeformed length $l$. The challenge in  computer modeling is the quantification of the friction and spring forces and the numerical integration of the equations of motion. 
 Dimensionless units are used so that $m=1$, $k=1$, and $l=50$. The value for $l$ was chosen for the sake of an easier 
graphical visualization (spring length corresponding to $50$ pixels). 
In order to have these dimensionless values to
correspond to a realistic experimental situation, the units are chosen as:  $[m]=0.1158$~kg for mass, $[k]=19.8$~N/m for spring 
constant, and $[l]=1.4\cdot 10^{-3}$~m for length. The units of the other quantities follow from 
dimensional considerations. The time, velocity and force units are thus, $[t]=\sqrt{[m]/[k]}=0.0765$~s, $[u]=[l]/[t]=0.0183$~m/s, and $
[F]=[k]\cdot [l]=0.0277$~N, respectively.

The equation of motion for the $i$-th block of the chain is
\begin{equation}
\ddot{x_i}=F_e(\Delta l_{-})-F_e(\Delta l_{+})+F_f \left( v_{r_i},F_e(\Delta l_{-})-F_e(\Delta l_{+}) \right),
  \label{eq:newton}
\end{equation}
where $\Delta l_{-}=x_i-x_{i-1}-50$ and $\Delta l_{+}=x_{i+1}-x_{i}-50$, respectively, and $v_{r_i}$ is the relative velocity of the block to the conveyor belt.
The elastic force, $F_e$, and the friction force, $F_f$, are defined below.  The numerical method used  for integrating
the equations of motions are presented in Appendix B, and the time-step was chosen as $dt=0.01$ 
time units. 

\begin{figure}[ht!]
  \centering
  \begin{center}
    \includegraphics[scale=0.7]{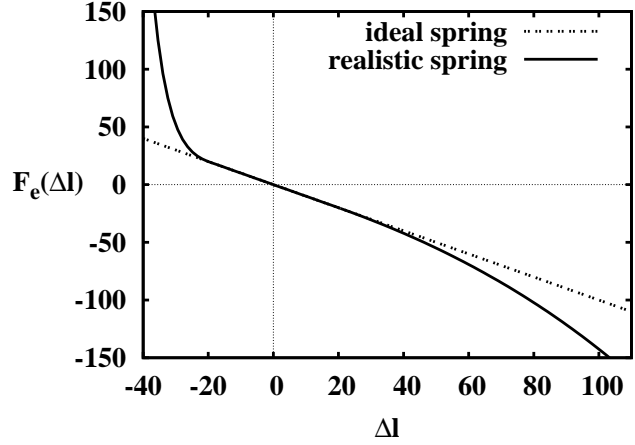}
  \end{center}
  \caption{The nonlinear spring force profile used in this study, in comparison with the ideal spring force profile.}
  \label{fig9}
\end{figure}

The elastic force $F_e$ in the springs is linear, up to a certain deformation value, $\Delta l_{max}$. For higher deformations, 
this dependence is assumed to become exponential, with an exponent bigger (in modulus) for 
negative deformations (see Fig. \ref{fig9}). Accordingly,

\begin{equation}
F_{e}(\Delta l) = 
{
\begin{cases}
  \Delta l_{max} + \frac{1}{b_1}e^{b_1(|\Delta l|-\Delta l_{max})}-\frac{1}{b_1}, \\  \hspace{3 cm} \text{if} \ \Delta l < -\Delta l_{max}, \\
  -\Delta l, 	\\ \hspace{3 cm} \text{if} \ -\Delta l_{max} \leqslant \Delta l \leqslant \Delta l_{max},\\
  -[\Delta l_{max}+\frac{1}{b_2}e^{b_2(|\Delta l|-\Delta l_{max})}-\frac{1}{b_2}], 	\\ \hspace{3 cm} \text{if} \  \Delta l > \Delta l_{max},
\end{cases}
}
\end{equation} 
where we have chosen $\Delta l_{max}=20$, $b_1=0.2$ for $\Delta l<0$ and $b_2=0.01$ for $\Delta l \geqslant 0$. With these choices the model is quite realistic for an experimental 
realization (see for example \cite{Bulcsu}) since the nonlinearity of spring forces is taken into account and collisions between blocks are avoided. At the same time the choice of parameters $b_1$ and $b_2$ is somewhat ad-hoc and we have not carried out a systematic change of them.

For the velocity dependent friction the simple  Coulomb's law of friction is used. 
Both the static and kinetic friction forces are independent of velocity modulus. 
A block remains in the stick state until the resultant external force 
$F_{ex}$ exceed the value of the static friction force, $F_{st}$. 
For higher external force values the block starts to slide in the presence of 
the kinetic friction force $F_{k}$. We assume that, the ratio of the static and kinetic friction forces 
$F_{k}/F_{st}=f_s$ is constant. The friction force $F_{f}$ acting on the block depends both 
on $S{(v_r)}$, where $S(\cdot)$ is the signum function 
and $v_r$ is the block's speed relative to the  conveyor belt, and on the value of the resultant 
external forces $F_{ex}$ acting on it. In our 1D setup, the friction force orientation 
is given only by its sign:

\begin{equation}
  F_{f}(v_r, F_{ex})=
  \begin{cases}
  -F_{ex},		& \text{if} \ v_r=0, |F_{ex}| < F_{st}, \\
  -S{(v_r)}f_{s}F_{st},	& \text{if} \ v_r\neq 0,
  \end{cases}
\end{equation} 
where $v_r=v-u$  and $v$ is the velocity of the block relative to the laboratory frame. 
In order to use the same friction force value as in our previous experiments \cite{Bulcsu}, 
in the dimensionless units the static friction 
force is taken as $ F_{st} =71.4$ and we considered $f_s=0.45$.

For reproducing also the empirically measured fat-tail (power law) distribution 
of the logarithmic returns, the speed of the conveyor belt was varied randomly 
 with a power law distribution, normalized in the $u \in [5,100]$ velocity interval. 
The exponent of the power function was chosen to be $\alpha=-3.0$.  In such way the model reproduces 
the empirically measured power law distributions for positive and also for negative returns 
\cite{Chakraborti}.

Assuming that the daily volatility of the logarithmic returns for DJIA and the one given by our model 
should be the same, we determined the $\delta t$ sampling time for 
the dynamics of $X(t)$. In the units of our simulations, we obtained  
$\delta t=5$ (500 integration steps). 
With this sampling frequency the volatility of log-returns  for the artificial index turns to be 
$\sigma_{a}=0.011$, which corresponds to the empirical value up to three decimals.

\section*{Appendix B - numerical integration of the equation of motions}
\label{sec:numerical_methods}
We briefly describe here the method used to integrate the equations of motion (\ref{eq:newton})
and to handle the discontinuous stick-slip dynamics of the blocks. 
When a block is stuck to the conveyor belt,
it moves together with it at constant velocity $u$. Therefore, the position 
of the $i$th block relative to the ground is calculated with the simple 
\begin{equation}
 x_i(t+dt)=x_i(t)+u\cdot dt,
\end{equation}
equation. When a block is slipping relative to the belt, the basic Verlet method
\begin{equation}
 x_i(t+dt)=2x_i(t)-x_i(t-dt)+a_i(t){dt}^2+O({dt}^4)
\end{equation}
is used to update its position. As can be seen, this is a third order method, which can be extended 
also to the velocity space as:
\begin{eqnarray}
 v_i(t+dt)=\frac{x_i(t)-x_i(t-dt)}{dt}+ \nonumber \\
 +\frac{1}{6}dt[11a_i(t)-2a_i(t-dt)] +O({dt}^3).
\end{eqnarray}
The instance when the $i$th block sticks to the belt is found when the relative velocity $v_{r_i}$ 
changes its sign, while a block starts to slip 
when the sum of external forces acting on it exceed the maximal value of 
static friction forces, i.e. when $F_{ex}-F_{st}>0$. 
A more complicated stochastic numerical method was also developed to handle the stick-slip dynamics, 
but was found not to significantly alter the presented results \cite{Bulcsu}.

\bibliographystyle{model1-num-names}

\end{document}